\documentclass[preprint,superscriptaddress,nofootinbib]{revtex4}
\usepackage[dvips]{graphicx}
\usepackage{amsmath,amssymb,color}

\newcommand{\red}[1]{{\color[rgb]{1,0,0} #1}}

\newcommand{\MNS}{{\text{MNS}}}

\newcommand{\eV}{{\text{eV}}}

\newcommand{\BR}{\text{BR}}
\newcommand{\Det}{\text{Det}}

\newcommand{\U}{{\text{U}}}
\newcommand{\SU}{{\text{SU}}}

\newcommand{\ellp}{\ell^\prime}

\begin{document}

\preprint{UT-HET 108}

\title{
Probing Models of Neutrino Masses
via the Flavor Structure of the Mass Matrix
}

\author{Shinya Kanemura}
\email{kanemu@sci.u-toyama.ac.jp}
\affiliation{
Department of Physics,
University of Toyama,
3190 Gofuku,
Toyama 930-8555, Japan
}
\author{Hiroaki Sugiyama}
\email{sugiyama@sci.u-toyama.ac.jp}
\affiliation{
Department of Physics,
University of Toyama,
3190 Gofuku,
Toyama 930-8555, Japan
}


\begin{abstract}
 We discuss what kinds of combinations of Yukawa interactions
can generate the Majorana neutrino mass matrix.
 We concentrate on
the flavor structure of the neutrino mass matrix
because it does not depend on details of the models
except for Yukawa interactions
while determination of the overall scale of the mass matrix
requires to specify also the scalar potential
and masses of new particles.
 Thus,
models to generate Majorana neutrino mass matrix
can be efficiently classified according to
the combination of Yukawa interactions.
 We first investigate the case
where Yukawa interactions with only leptons are utilized.
 Next, we consider
the case with Yukawa interactions
between leptons and gauge singlet fermions,
which have the odd parity under the unbroken $Z_2$ symmetry.
 We show that combinations of Yukawa interactions for these cases
can be classified into only three groups.
 Our classification would be
useful for the efficient discrimination of models
via experimental tests
for not each model but just three groups of models.
\end{abstract}

\maketitle


\section{Introduction}

 Thanks to the discovery of a Higgs boson $h$
at the CERN Large Hadron Collider~(LHC)~\cite{ref:2012Jul},
we have entered the era to explore the origin of particle masses.
 Coupling constants of $W^\pm$, $Z$, $t$, $b$, and $\tau$ with $h$
are measured at the LHC~\cite{ref:combined},
and they are consistent with predicted values in the Standard Model~(SM).
 These results strongly suggest that
masses of gauge bosons and charged fermions
are generated by the vacuum expectation value
of the Higgs field, which provides $h$,
as predicted in the SM\@.
 Thus,
the mechanism to generate their masses in the SM
was confirmed.
 On the other hand,
neutrino masses are not included in the SM
although
neutrino oscillation data uncovered
that neutrinos have their masses~\cite{ref:SK-atm, ref:SNO}.
 It is easy to add neutrino mass terms
$m_\nu^{} \overline{\nu_L^{}} \nu_R^{}$ to the SM
similarly to the other fermion mass terms
by introducing right-handed neutrinos $\nu_R^{}$.
 However,
since the neutrino is a neutral fermion
in contrast to the other fermions in the SM,
another possibility of its mass term exists.
 That is the Majorana mass term,
$(1/2) m_\nu^{} \overline{\nu_L^{}} (\nu_L^{})^c$.
 This unique possibility
could be the reason why neutrinos
are much lighter than the other fermions.
 New physics models for the Majorana neutrino mass
can be found in e.g.\ Refs.%
~\cite{Zee:1980ai, Wolfenstein:1980sy,
Zee:1985id, ref:ZB-B,
Cheng:1980qt,
ref:GNR,
ref:HTM,
ref:KNT,
ref:AKS,
ref:Ma,
ref:seesaw,
Foot:1988aq,
Mohapatra:1986bd, Khalil:2010iu, Fraser:2014yha,
Pilaftsis:1991ug, Dev:2012sg, Barr:2003nn, Wang:2015saa,
Kanemura:2011mw, Kanemura:2014rpa, Kanemura:2012rj,
Kubo:2006rm, Kanemura:2011vm, Aoki:2013gzs, Dasgupta:2013cwa,
Ma:2008cu, Kanemura:2010bq, Aoki:2011yk, Kumericki:2012bf, Okada:2012np,
Brdar:2013iea, Aranda:2015xoa, 
Ahriche:2014cda, Ahriche:2014oda, Chen:2014ska, Ahriche:2015wha,
Culjak:2015qja,
Lindner:2011it, Okada:2014qsa, Hatanaka:2014tba,
Kajiyama:2013zla, Restrepo:2015ura,
McDonald:2013kca, Law:2013gma, Law:2013saa,
Kajiyama:2013rla, Okada:2014nsa, Okada:2014oda, Okada:2015bxa,
Kashiwase:2015pra, Nishiwaki:2015iqa, Okada:2015hia}.

 The overall scale of
the neutrino mass matrix $m_\nu$
generated in new physics models
is determined by
the structure (tree level, one-loop level, and so on)
of the diagram to generate $m_\nu$,
masses of new particles in the diagram
and coupling constants in the diagram.
 This means that
the determination of the overall scale of $m_\nu$
requires to specify many parts
of the Lagrangian of each model.
 On the other hand,
the flavor structure (ratios of elements) of $m_\nu$
is simply determined by
the product of Yukawa coupling matrices
and fermion masses.
 Thus,
models to generate $m_\nu$
can efficiently be classified
according to the combination of Yukawa coupling matrices
and fermion masses
without the detail of these models.
 When we construct a new model to generate neutrino masses,
it will be noticed indeed that
the flavor structure is the key to find
an appropriate set of model parameters
although the overall scale of $m_\nu$
can be easily tuned by using some parameters
in the scalar potential.

 In this letter,
we first classify models for Majorana neutrino masses
according to combination of Yukawa interaction
between leptons without introducing new fermions.
 Next,
we do the classification
for the case where
gauge singlet fermions are introduced
such that they have the odd parity
under the unbroken $Z_2$ symmetry
which can be utilized to stabilize the dark matter.
 For Yukawa interactions of these new fermions with leptons,
$Z_2$-odd scalars are also introduced.
 We find that
models can be classified into only three groups.
 The classification
could be useful to approach
efficiently the origin of Majorana neutrino masses
with experimental tests of
not each model but each group of models.

 Models of neutrino masses
can also be classified according to
topologies of diagrams~\cite{ref:diagram}
or decompositions of higher mass-dimensional operators%
~\cite{ref:higher-dim}.
 They seem useful to find new models
and increase the number of models
in order to exhaust all possibilities.
 In contrast with these classifications,
ours would be useful
to simplify the situation where many models exist.



\section{Classification of Flavor Structure}

\begin{table}[t]
\begin{tabular}{c||c|c|c|c|c}
Scalar
 & $\SU(2)_L$
 & $\U(1)_Y$
 & L\#
 & Yukawa
 & Note
\\
\hline
\hline
$s_1^+$
 & ${\bf\underline{1}}$
 & $1$
 & $-2$
 & $
    (Y_A^s)_{\ell\ellp}
     \Bigl[
      \overline{L_\ell^{}}\, \epsilon L_{\ell^\prime}^c\, s_1^-
     \Bigr]
   $
 & Antisymmetric
\\[1mm]
\hline
$s^{++}$
 & ${\bf\underline{1}}$
 & $2$
 & $-2$
 & $
    (Y_S^s)_{\ell\ellp}
     \Bigl[
      \overline{(\ell_R^{})^c}\, \ell^\prime_R\, s^{++}
     \Bigr]
   $
 & Symmetric
\\[1mm]
\hline
$
\Phi_2
=
\begin{pmatrix}
 \phi_2^+\\
 \phi_2^0
\end{pmatrix}
$
 & ${\bf\underline{2}}$
 & $\displaystyle\frac{1}{\,2\,}$
 & $0$
 & $
    y_\ell^{}
    \Bigl[
     \overline{L_\ell^{}}\, \Phi_2\, \ell_R^{}
    \Bigr]
   $
 & Diagonal
\\
\hline
$
\Delta
=
\begin{pmatrix}
 \displaystyle
 \frac{\ \ \Delta^+}{\sqrt{2}}
  & \Delta^{++}\\[3mm]
 \Delta^0
  & \displaystyle -\frac{\ \ \Delta^+}{\sqrt{2}}
\end{pmatrix}
$
 & ${\bf\underline{3}}$
 & $1$
 &$-2$
 & $
    (Y_S^\Delta)_{\ell\ell^\prime}
    \Bigl[
     \overline{L_\ell^{}}\, \Delta^\dagger \epsilon\,
     L_{\ell^\prime}^c
    \Bigr]
   $
 & Symmetric
\end{tabular}
\caption
{
 Scalar bosons
which can have Yukawa interactions with leptons
without introducing new fermions.
 The Yukawa matrix $Y_A$ is antisymmetric,
while $Y_S^s$ and $Y_S^\Delta$ are symmetric.
 The lepton number~(L\#) is assigned
to each of scalar fields
such that the Yukawa interactions conserve the L\#
as a convention.
 Then, the L\# is broken in the scalar potential.
}
\label{tab:particle-I}
\vspace*{5mm}
\end{table}

\begin{figure}[t]
\begin{center}
\includegraphics[scale=0.7]{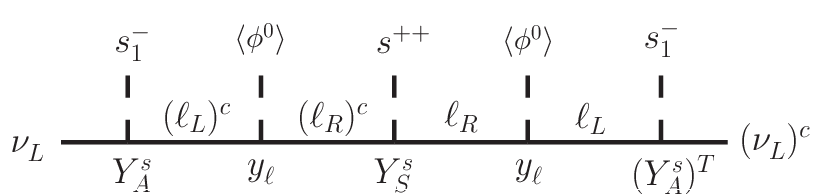}
\vspace*{-5mm}
\caption{
 The diagram of the fermion line
for the combination in eq.~\eqref{eq:mass-ZB}.
}
\label{fig:ZB-type}
\end{center}
\end{figure}

\begin{figure}[t]
\begin{center}
\includegraphics[scale=0.7]{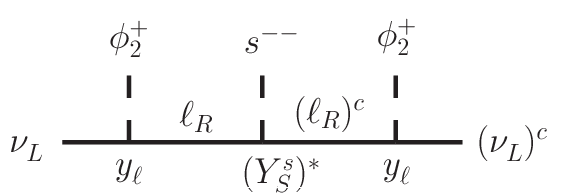}
\vspace*{-5mm}
\caption{
 The diagram of the fermion line
for the combination in eq.~\eqref{eq:mass-CL}.
}
\label{fig:CL-type}
\end{center}
\end{figure}

\begin{figure}[t]
\begin{center}
\includegraphics[scale=0.7]{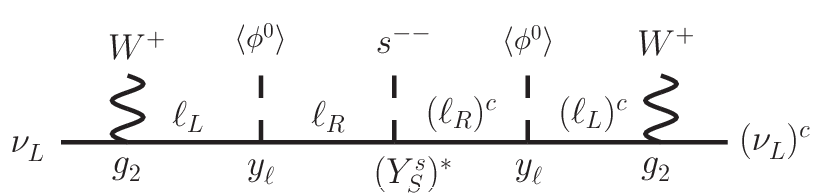}
\vspace*{-5mm}
\caption{
 The diagram of the fermion line
for the combination in eq.~\eqref{eq:mass-GNR}.
}
\label{fig:GNR-type}
\end{center}
\end{figure}

\begin{figure}[t]
 \begin{minipage}{0.45\hsize}
  \vspace*{1mm}
  \begin{center}
   \includegraphics[scale=0.7]{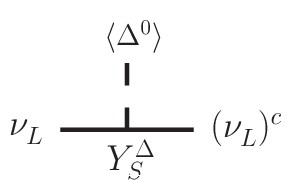}
   \vspace*{-5mm}
   \caption{
    The diagram of the fermion line
    for the combination in eq.~\eqref{eq:mass-HTM}.
   }
   \label{fig:HTM-type}
  \end{center}
 \end{minipage}
\hspace*{8mm}
 \begin{minipage}{0.45\hsize}
  \begin{center}
   \includegraphics[scale=0.7]{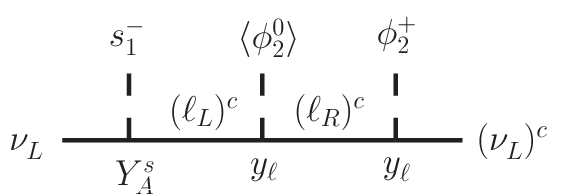}
   \vspace*{-5mm}
   \caption{
    The diagram of the fermion line
    for the combination in eq.~\eqref{eq:mass-ZW}.
   }
   \label{fig:ZW-type}
  \end{center}
 \end{minipage}
\vspace*{5mm}
\end{figure}

 First,
we introduce only scalar fields
listed in Table~\ref{tab:particle-I},
which have Yukawa interactions with leptons.
 We do not always introduce all of them,
and we utilize only scalar bosons
for required Yukawa interactions.
 For the Yukawa interaction
with the second $\SU(2)_L$-doublet scalar field $\Phi_2$,
the flavor changing neutral current
is forbidden by utilizing a softly-broken $Z_2$ symmetry
as usually done in the two Higgs doublet models.
 In order to obtain $m_\nu^{}$,
we try to connect $\nu_L^{}$ to $(\nu_L^{})^c$
by using these Yukawa interactions and the weak interaction.
 We do not care how scalar lines are closed
because we concentrate on the flavor structure of $m_\nu^{}$.
 Each charged lepton%
~($\ell_L$, $\ell_R$, $(\ell_L)^c$, $(\ell_R)^c$)
should appear only once on the fermion line
in order to obtain the simplest combinations,
which would give the largest contribution to $m_\nu^{}$.
 In addition,
$\ell_L$ and $\ell_R$ should appear
only in the next to each other
on the fermion line.
 If they do not,
the replacement of the structure between them
with the mass term of $\ell$
can give the simpler combination%
\footnote{
 Although the electron Yukawa coupling is small,
the diagonal matrix $y_\ell$ would not be negligible
because of the tau Yukawa coupling.
}.
 It is assumed that
$m_\nu^{}$ is generated via a solo mechanism
(a solo kind of fermion lines).
 Then,
we find that only the following five combinations%
\footnote
{
 Notice that
another possible combination
$Y_A^s\, g_2 + (Y_A^s\, g_2)^T$
becomes zero.
}
connect $\nu_L^{}$ and $(\nu_L^{})^c$:
\begin{eqnarray}
m_\nu^{}
 &\propto& Y_A^s\, y_\ell^{}\, Y_S^s\, y_\ell^{}\, (Y_A^s)^T,
\label{eq:mass-ZB}
\\
%
%
m_\nu^{}
 &\propto& y_\ell^{}\, (Y_S^s)^\ast\, y_\ell^{},
\label{eq:mass-CL}
\\
%
%
m_\nu^{}
 &\propto& g_2^{}\, y_\ell^{}\, (Y_S^s)^\ast\, y_\ell^{}\, g_2^{},
\label{eq:mass-GNR}
\\
%
%
m_\nu^{}
 &\propto& Y_S^\Delta,
\label{eq:mass-HTM}
\\
%
%
m_\nu^{}
 &\propto& Y_A^s\, y_\ell^2 + (Y_A^s\, y_\ell^2)^T,
\label{eq:mass-ZW}
\end{eqnarray}
where Yukawa matrices $Y_A$, $Y_S^s$, $y_\ell$,
and $Y_S^\Delta$ are defined in Table~\ref{tab:particle-I}.
 Diagrams of fermion lines
for combinations in eqs.~\eqref{eq:mass-ZB}-\eqref{eq:mass-ZW}
are shown in Figs.~\ref{fig:ZB-type}-\ref{fig:ZW-type},
respectively.
 The $\SU(2)_L$ gauge coupling constant $g_2$
is shown for clarity
although the weak interaction is flavor blind.
 The combination in eq.~\eqref{eq:mass-GNR}
gives at least a dimension-9 operator
for the Majorana neutrino mass
while the others can be a dimension-5 one.

 The combination in eq.~\eqref{eq:mass-ZW}
is the one in the Zee-Wolfenstein model%
~\cite{Zee:1980ai, Wolfenstein:1980sy}
of the Majorana neutrino mass at the one-loop level,
which has been excluded already
by the neutrino oscillation data~\cite{He:2003ih}.
 Thus,
this combination is ignored below.
 An example for $m_\nu^{}$ in eq.~\eqref{eq:mass-ZB}
is the Zee-Babu~(ZB) model~\cite{Zee:1985id, ref:ZB-B},
which generates $m_\nu^{}$ at the two-loop level.
 The structure in eq.~\eqref{eq:mass-CL} is given
in a model in Ref.~\cite{Cheng:1980qt}
by Cheng and Li~(the CL model),
which also generates $m_\nu^{}$ at the two-loop level%
\footnote
{
 In Ref.~\cite{Cheng:1980qt},
scalar lines of $\phi_2^+$ and $s^{--}$
are closed in a little bit complicated way.
 Instead of that,
it seems the simplest to introduce
an $\SU(2)_L$-doublet scalar field with
the hypercharge $Y=3/2$.
}.
 The Gustafsson-No-Rivera~(GNR) model~\cite{ref:GNR}
is an example for the combination in eq.~\eqref{eq:mass-GNR},
in which $m_\nu^{}$ is generated at the tree-loop level.
 Scalar lines of $W^+$ and $s^{--}$ are connected
at the one-loop level by introducing
the unbroken $Z_2$ symmetry and $Z_2$-odd scalar fields,
which provide a dark matter candidate.
 The structure in eq.~\eqref{eq:mass-HTM}
is given at the tree level,
and an example is the Higgs triplet model~(HTM)~\cite{Cheng:1980qt, ref:HTM}.
 Since eqs.~\eqref{eq:mass-CL} and \eqref{eq:mass-GNR}
have the same flavor structure,
that of $m_\nu^{}$
is given by only three combinations
of Yukawa matrices:
$Y_A^s\, y_\ell^{}\, Y_S^s\, y_\ell^{}\, (Y_A^s)^T$,
$y_\ell^{}\, (Y_S^s)^\ast\, y_\ell^{}$,
and $Y_S^\Delta$.

\begin{table}[t]
\begin{tabular}{c||c|c|c|c}
Scalar
 & $\SU(2)_L$
 & $\U(1)_Y$
 & Yukawa
 & Note
\\
\hline
\hline
$\red{s_2^+}$
 & ${\bf\underline{1}}$
 & $1$
 & $
    Y_{\ell i}^s
    \Bigl[
     \overline{(\ell_R^{})^c}\, \red{\psi_{iR}^0}\, \red{s_2^+}
    \Bigr]
   $
 & Arbitrary
\\[1mm]
\hline
$
\red{\eta}
=
\begin{pmatrix}
 \red{\eta^+}\\
 \red{\eta^0}
\end{pmatrix}
$
 & ${\bf\underline{2}}$
 & $\displaystyle\frac{1}{\,2\,}$
 & $
    Y_{\ell i}^\eta
    \Bigl[
     \overline{L_\ell}\, \epsilon\, \red{\eta^\ast}\, \red{\psi_{iR}^0}
    \Bigr]
   $
 & Arbitrary
\end{tabular}
\caption
{ 
 Scalar bosons
for Yukawa interactions of
gauge singlet fermion $\red{\psi_{iR}^0}$ with leptons.
 These scalar bosons and $\psi_{iR}^0$ are $Z_2$-odd fields.
 Structures of Yukawa matrices $Y^s$ and $Y^\eta$
are arbitrary.
 When $\psi_R^0$ has $\text{L\#} = x$,
lepton numbers $-x-1$ and $x-1$ are
assigned to $s_2^+$ and $\eta$, respectively,
such that their Yukawa interactions
conserve the L\# as a convention.
 The L\# is broken in the scalar potential
and/or $M_\psi$.
}
\label{tab:particle-II}
\vspace*{5mm}
\end{table}

\begin{figure}[t]
\begin{center}
\includegraphics[scale=0.7]{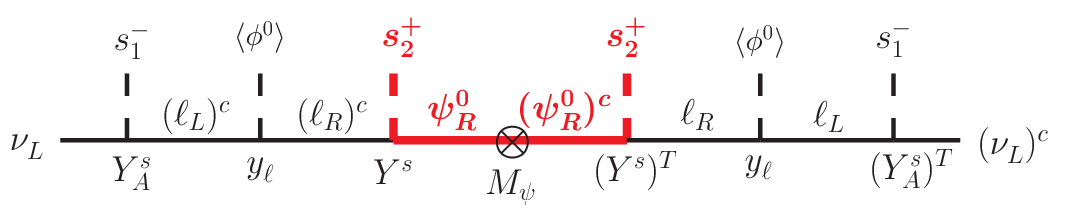}
\vspace*{-5mm}
\caption{
 The diagram of the fermion line
for the combination in eq.~\eqref{eq:mass-KNT}.
 Bold red lines are for the $Z_2$-odd particles.
}
\label{fig:KNT-type}
\end{center}
\end{figure}

\begin{figure}[t]
\begin{center}
\includegraphics[scale=0.7]{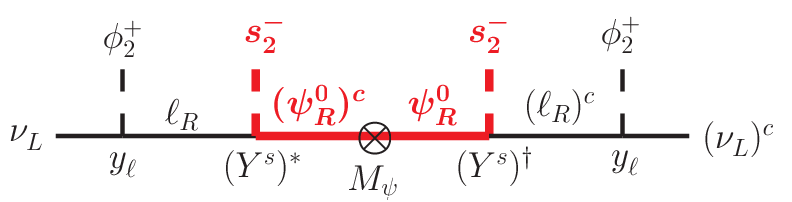}
\vspace*{-5mm}
\caption{
 The diagram of the fermion line
for the combination in eq.~\eqref{eq:mass-AKS}.
 Bold red lines are for the $Z_2$-odd particles.
}
\label{fig:AKS-type}
\end{center}
\end{figure}

\begin{figure}[t]
\begin{center}
\includegraphics[scale=0.7]{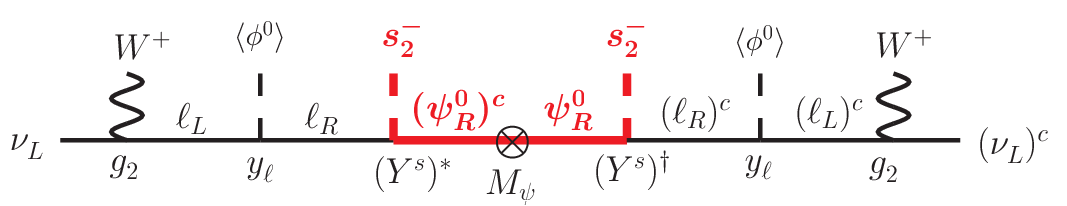}
\vspace*{-5mm}
\caption{
 The diagram of the fermion line
for the combination in eq.~\eqref{eq:mass-GNR-Z2}.
 Bold red lines are for the $Z_2$-odd particles.
}
\label{fig:GNR-Z2-type}
\end{center}
\end{figure}

\begin{figure}[t]
\begin{center}
\includegraphics[scale=0.7]{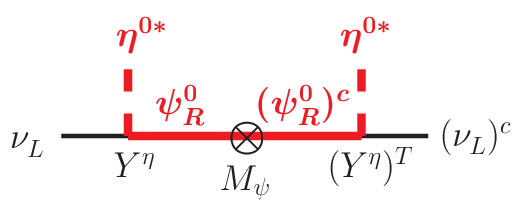}
\vspace*{-5mm}
\caption{
 The diagram of the fermion line
for the combination in eq.~\eqref{eq:mass-Ma}.
 Bold red lines are for the $Z_2$-odd particles.
}
\label{fig:Ma-type}
\end{center}
\end{figure}

Next,
we impose the unbroken $Z_2$ symmetry to models
and introduce gauge singlet fermions $\psi_{iR}^0$
as the $Z_2$-odd fields.
 The fermions have Majorana mass terms,
$(1/2) M_{\psi i} \overline{(\psi_{iR}^0)^c} \psi_{iR}^0$.
 We can take the basis where $M_\psi$ is diagonalized
without loss of generality.
 For Yukawa interactions of $\psi_{iR}^0$ with leptons,
scalar fields in Table~\ref{tab:particle-II}
are also introduced as $Z_2$-odd fields.
 Scalar fields in Table~\ref{tab:particle-I} and the SM fields
are $Z_2$-even ones.
 Then,
the lightest $Z_2$-odd particle becomes stable.
 If the lightest $Z_2$-odd particle is neutral one,
it can be a dark matter candidate.
 We find that
the Majorana neutrino mass matrix can be obtained
by the following four kinds of combinations
of Yukawa matrices and the weak interaction
in addition to the five combinations
in eqs.~\eqref{eq:mass-ZB}-\eqref{eq:mass-ZW}:
\begin{eqnarray}
m_\nu^{}
 &\propto&
 Y_A^s\, y_\ell^{}\, Y^s\, M_\psi^{-1}\, (Y^s)^T\, y_\ell^{}\, (Y_A^s)^T,
\label{eq:mass-KNT}
\\
%
%
m_\nu^{}
 &\propto&
 y_\ell^{}\, (Y^s)^\ast\, M_\psi^{-1}\, (Y^s)^\dagger\, y_\ell^{},
\label{eq:mass-AKS}
\\
%
%
m_\nu^{}
 &\propto&
 g_2^{}\, y_\ell^{}\, (Y^s)^\ast\, M_\psi^{-1}\, (Y^s)^\dagger\, y_\ell^{}\, g_2^{},
\label{eq:mass-GNR-Z2}
\\
%
%
m_\nu^{}
 &\propto& Y^\eta\, M_\psi^{-1}\, (Y^\eta)^T,
\label{eq:mass-Ma}
\end{eqnarray}
where Yukawa matrices $Y^s$ and $Y^\eta$
are defined in Table~\ref{tab:particle-II}.
 Figures~\ref{fig:KNT-type}-\ref{fig:Ma-type}
correspond to diagrams of fermion lines
for combinations in eqs.~\eqref{eq:mass-KNT}-\eqref{eq:mass-Ma},
respectively.
 The part $M_\psi^{-1}$ is given by
assuming $\psi_{iR}^0$ are heavier than the other particles.
 If it is not the case,
$M_\psi^{-1}$ can be replaced with $M_\psi$.

 The Krauss-Nasri-Trodden~(KNT) model~\cite{ref:KNT}
of $m_\nu$ at the three-loop level
is an example for the combination in eq.~\eqref{eq:mass-KNT}.
 The structure in eq.~\eqref{eq:mass-AKS}
is realized, for example,
in the Aoki-Kanemura-Seto~(AKS) model~\cite{ref:AKS}
at the three-loop level
by introducing the $Z_2$-odd real singlet scalar boson.
 Since the three-loop diagram
utilizes the scalar interaction with two Higgs doublet fields,
the AKS model can explain not only $m_\nu$ and the dark matter
but also the baryon asymmetry of the universe
via the electroweak baryogenesis scenario.
 An example of the combination in eq.~\eqref{eq:mass-Ma}
is the Ma model~\cite{ref:Ma},
where $m_\nu$ is generated at the one-loop level.
 No model is known for $m_\nu^{}$ in eq.~\eqref{eq:mass-GNR-Z2}%
\footnote
{
 The combination in eq.~\eqref{eq:mass-GNR-Z2}
gives at the least a dimension-9 operator for $m_\nu^{}$,
and it might be four-loop realization at the least.
 Then, too small neutrino masses
might be generated.
}.
 Flavor structures of combinations
in eqs.~\eqref{eq:mass-AKS} and \eqref{eq:mass-GNR-Z2}
are the same
because the weak interaction does not change the flavor.
 Therefore,
the flavor structure of $m_\nu^{}$
is determined by three combinations
when we use the Yukawa interactions in Table~\ref{tab:particle-II}:
$Y_A^s\, y_\ell^{}\, Y^s\, M_\psi^{-1}\, (Y^s)^T\, y_\ell^{}\, (Y_A^s)^T$,
$y_\ell^{}\, (Y^s)^\ast\, M_\psi^{-1}\, (Y^s)^\dagger\, y_\ell^{}$,
and $Y^\eta\, M_\psi^{-1}\, (Y^\eta)^T$.

 It is clear that
combinations in eqs.~\eqref{eq:mass-ZB}-\eqref{eq:mass-HTM}
and eqs.~\eqref{eq:mass-KNT}-\eqref{eq:mass-Ma}
can be classified further to only the following three groups:
\begin{eqnarray}
\text{Group-I} : \quad
m_\nu^{}
 &\propto&
 Y_A^s\, y_\ell^{}\, X_{SR}\, y_\ell^{}\, (Y_A^s)^T,
\label{eq:mass-I}
\\
%
%
\text{Group-II} : \quad
m_\nu^{}
 &\propto&
 y_\ell^{}\, X_{SR}^\ast\, y_\ell^{},
\label{eq:mass-II}
\\
%
%
\text{Group-III} : \quad
m_\nu^{}
 &\propto& X_{SL},
\label{eq:mass-III}
\end{eqnarray}
where symmetric matrices
$X_{SR}$ and $X_{SL}$ are given by
\begin{eqnarray}
X_{SR}
&=&
 Y_S^s, \quad
 Y^s M_\psi^{-1} (Y^s)^T , \quad
 Y^s M_\psi (Y^s)^T ,
\label{eq:XSR}
\\
%
%
X_{SL}
&=&
 Y_S^\Delta, \quad
 Y^\eta M_\psi^{-1} (Y^\eta)^T , \quad
 Y^\eta M_\psi (Y^\eta)^T .
\label{eq:XSL}
\end{eqnarray}
 The matrix $X_{SR}$ is
for the effective interactions
of right-handed charged leptons
while the matrix $X_{SL}$ is
for the ones of left-handed leptons.
 As long as we concentrate on the flavor structure,
it seems difficult to discriminate
the origin of $X_{SR}$~($X_{SL}$)
in eq.~\eqref{eq:XSR}~(eq.~\eqref{eq:XSL}).

 We mention here
the type-I~\cite{ref:seesaw}
and the type-III seesaw~\cite{Foot:1988aq} models,
where gauge singlet fermions~(for the type-I)
or $\SU(2)_L$-triplet Majorana fermions~(for the type-III)
are introduced.
 The structure of $m_\nu^{}$ in these models
can be included in the Group-III
because Yukawa matrices $Y_A$ and $y_\ell$ are not used
to generate $m_\nu^{}$.
 However,
they are exceptions
because new scalar fields are not introduced.
 Discussion in the next section
(namely, $\tau \to \overline{\ell}_1 \ell_2 \ell_3$%
~$(\ell_1, \ell_2, \ell_3 = e, \mu)$
for the Group-III)
is not applicable for these models%
\footnote
{
 There is the box diagram with the $W$ boson and neutral fermions
from $\SU(2)_L$-singlet or triplet,
but the interaction of the neutral fermions with $W$
is suppressed by $\sqrt{m_\nu^{}/M_R}$%
~(the mixing between $\nu_L$ and the fermions),
where $M_R$ denotes the fermion mass.
}.



\section{Discussion}

 The neutrino mass matrix $m_\nu^{}$ is expressed as
$U_\MNS^\ast
\text{diag}(m_1 e^{i\alpha_{12}}, m_2, m_3 e^{i\alpha_{32}})
U_\MNS^\dagger$,
where $m_i$~($i=1\text{-}3$) are the neutrino mass eigenvalues,
$\alpha_{12}$ and $\alpha_{32}$ are the Majorana phases~\cite{ref:Mphase},
and $U_\MNS$ is the Maki-Nakagawa-Sakata~(MNS) matrix~\cite{Maki:1962mu}
of the lepton flavor mixing.
 The Group-I gives $m_1 = 0$ or $m_3 = 0$
because of $\Det(m_\nu) \propto \Det(Y_A) = 0$.
 Although this has been known
for the Zee-Babu model~\cite{ref:ZB-B}
(an example of models in the Group-I),
our statement is more model-independent.
 The Group-I is excluded
if the absolute neutrino mass is directly measured
at the KATRIN experiment~\cite{Osipowicz:2001sq}
whose estimated sensitivity is $0.35\,\eV$
at $5\,\sigma$ confidence level.
 The indirect bound on the sum of neutrino masses,
$\sum_i m_i < 0.23\,\eV$~(90\% confidence level),
was obtained by cosmological observations~\cite{Ade:2015xua},
and sensitivity to $\sum_i m_i = {\mathcal O}(0.01)\,\eV$
is expected in future experiments~\cite{Abazajian:2013oma}.

 The flavor structure of $m_\nu$ is
constrained by the neutrino oscillation data,
and the constrained structure can be translated into
constraints on the flavor structure~(ratios of elements) of
$X_{SR}$ of the Group-II and $X_{SL}$ of the Group-III\@.
 Hereafter,
we denote $X_{SR}$ of the Group-II
and $X_{SL}$ of the Group-III as $X$ for simplicity.
 These interactions can cause
the lepton flavor violating~(LFV) decays
$\tau \to \overline{\ell}_1 \ell_2 \ell_3$%
~$(\ell_1, \ell_2, \ell_3 = e, \mu)$.
 Ratios of the decay branching ratios~(BR) of these LFV decays
can be determined by the flavor structure of $X$
independently on the overall scale of $m_\nu^{}$.
 In order to evade the strong constraint
$\BR(\mu \to \overline{e}ee) < 1.0 \times 10^{-12}$~\cite{Bellgardt:1987du},
LFV decays $\tau \to \overline{\ell}_1 \ell_2 \ell_3$
can be observed at the Belle~II experiment~\cite{Abe:2010gxa}
only for $X_{ee} = 0$ or $X_{e\mu} = 0$,
which constrains ratios of $\BR(\tau \to \overline{\ell}_1 \ell_2 \ell_3)$
as discussed in the HTM~(included in the Group-III)~\cite{ref:HTM-tlll}.
 For $X_{ee} = 0$~($X_{e\mu} = 0$),
LFV decays $\tau \to \overline{\ell} ee$
($\tau \to \overline{\ell} e\mu$) do not occur.
 Since $X_{e\ell}$ elements for the Group-II
are enhanced by $1/m_e$ for a given $m_\nu^{}$,
it is likely that
$\BR(\tau \to \overline{e}e\mu)$ for $X_{ee} = 0$
or $\BR(\tau \to \overline{e}ee)$ for $X_{e\mu} = 0$
is larger than the others.
 For $X_{ee} = X_{e\mu} = 0$,
only $\tau \to \overline{e} \mu\mu$ can be observed
for the Group-II as shown in the GNR model~\cite{ref:GNR},
while $\tau \to \overline{\mu} \mu\mu$ is also possible for the Group-III\@.
 Notice that $X_{ee} = 0$ for the Group-II and III
results in $(m_\nu^{})_{ee}=0$,
which is excluded
if the neutrinoless double beta decay%
~(See e.g.\ Ref.~\cite{Vergados:2012xy}) is observed
or $m_3 < m_1$~(the inverted mass ordering of neutrinos)
is determined by neutrino oscillation experiments%
~(See e.g.\ Ref.~\cite{Blennow:2013oma}).
 Notice also that
$(X_{SR})_{ee} = 0$ for the Group-I
does not mean $(m_\nu^{})_{ee} = 0$.
 Therefore,
if $(m_\nu^{})_{ee}=0$ is excluded
by these neutrino experiments,
the observation of $\tau \to \overline{\ell}ee$
indicates the Group-I
because the situation is inconsistent
for the Group-II and III\@.

 The discussion above did not require
the discovery of new particles.
 If a charged scalar boson is discovered
and dominantly decays into leptons,
the branching ratios are expected to be given by $Y_A$~($y_\ell$)
when the Group-I~(II) is assumed.
 The flavor structure of $y_\ell$ is known,
and decays via the $y_\ell$ are dominated by
the decay into $\tau$.
 The flavor structure of $Y_A$ is determined by
the neutrino oscillation data as
$(Y_A)_{e\mu}/(Y_A)_{e\tau} =
 - (U_\MNS)_{\tau 1}^\ast/(U_\MNS)_{\mu 1}^\ast$
and
$(Y_A)_{\mu\tau}/(Y_A)_{e\tau} =
 - (U_\MNS)_{e 1}/(U_\MNS)_{\mu 1}^\ast$
for $m_1 < m_3$.
 For $m_1 > m_3$,
they are given by
$(Y_A)_{e\mu}/(Y_A)_{e\tau} =
 - (U_\MNS)_{\tau 3}/(U_\MNS)_{\mu 3}$
and
$(Y_A)_{\mu\tau}/(Y_A)_{e\tau} =
 - (U_\MNS)_{e 3}^\ast/(U_\MNS)_{\mu 3}^\ast$.
 Ratios of decay branching ratios
$\BR(s_1^- \to e\nu) : \BR(s_1^- \to \mu\nu) : \BR(s_1^- \to \tau\nu)$
are roughly given by
$2 : 5 : 5$ for $m_1 < m_3$
and $2 : 1 : 1$ for $m_1 > m_3$~\cite{ref:ZB-pheno}.
 Therefore,
Group-I and II can be tested
by measuring leptonic decays of the charged scalar boson
at the collider experiments.

 When a group of models is favored
by the experiments discussed above,
we will try to discriminate models in the group
by using details of each model.
 For example,
the doubly-charged scalar boson
is introduced in the ZB model in the Group-I
while it does not exist in the KNT model of the Group-I\@.
 Thus,
if the doubly-charged scalar boson
is discovered at the collider experiments,
the ZB model would be favored among models in the Group-I\@.
 This is the same for the CL model and the GNR model in the Group-II
and the HTM in the Group-III\@.
 Even if groups of models have not been discriminated,
collider experiments can test each models
by measuring properties~(e.g.\ decay patterns) of new particles
as usually studied for model by model.



\section{Conclusion}

 In this letter,
we have studied the systematic classification of models
for generating Majorana neutrino masses $m_\nu^{}$
according to combinations of Yukawa interactions.
 If we use Yukawa interactions for leptons
by introducing new scalar fields relevant for these Yukawa interactions,
the flavor structure of $m_\nu$ is given by
three combinations:
$Y_A^s\, y_\ell^{}\, Y_S^s\, y_\ell^{}\, (Y_A^s)^T$,
$y_\ell^{}\, (Y_S^s)^\ast\, y_\ell^{}$,
and $Y_S^\Delta$.
 The Yukawa matrix $Y_A$ is antisymmetric
while $Y_S^s$ and $Y_S^\Delta$ are symmetric.
 The Yukawa couplings $y_\ell$
are proportional to charged lepton masses.
 For the case
where gauge singlet $Z_2$-odd fermions $\psi_{iR}^0$
and $Z_2$-odd scalar fields
are additionally introduced,
the flavor structure of $m_\nu$ is determined also by
$Y_A^s\, y_\ell^{}\, Y^s\, M_\psi^{-1}\, (Y^s)^T\, y_\ell^{}\, (Y_A^s)^T$,
$y_\ell^{}\, (Y^s)^\ast\, M_\psi^{-1}\, (Y^s)^\dagger\, y_\ell^{}$,
and $Y^\eta\, M_\psi^{-1}\, (Y^\eta)^T$.
 The Yukawa matrices $Y_S^s$ and $Y_S^\eta$ are symmetric,
and $M_\psi$ is the Majorana mass matrix for $\psi_{iR}^0$.
 Combining these results,
we have found that
models can be classified into only three groups:
$m_\nu^{} \propto
Y_A^s\, y_\ell^{}\, X_{SR}\, y_\ell^{}\, (Y_A^s)^T$,
$y_\ell^{}\, X_{SR}^\ast\, y_\ell^{}$,
and $X_{SL}$.
Here,
$X_{SR}$ and $X_{SL}$ are some symmetric matrices.
 Although the structure of $m_\nu^{}$
in the type-I seesaw and the type-III seesaw models
can be classified in the Group-III,
these models are exceptions to the discussion in this letter.
 Our classification enable us
to approach efficiently to the origin of Majorana neutrino masses
by testing not each model but each groups of models.

 We concentrated on Majorana neutrino masses
in this letter.
 The similar classification of models
for Dirac neutrino masses
is also desired because the nature may respect
the lepton number conservation.
 This will be presented elsewhere~\cite{ref:future}.

\begin{acknowledgments}
 This work was supported, in part,
by Grant-in-Aid for Scientific Research No.~23104006~(SK)
and Grand H2020-MSCA-RISE-2014 no.~645722
(Non Minimal Higgs) (SK).
\end{acknowledgments}

\end{document}